\documentclass[licencjacka,en]{pracamgr}
\usepackage{dirtytalk}
\usepackage{hyperref}
\usepackage{graphicx}
\usepackage{longtable}
\usepackage{cellspace}
\newcommand\Tspace{\rule{0pt}{2.6ex}}         
\newcommand\Bspace{\rule[-0.9ex]{0pt}{0pt}}   
\graphicspath{ {./images/} }

\autor{Tomasz Nowak}{429575}
\autori{Michał Staniewski}{429598}
\autorii{Mieszko Grodzicki}{429132}
\autoriii{Bartosz Smolarczyk}{429594}
\opiekun{mgr Michał Możdżonek\\
  Institute of Informatics\\
  }
\date{May 2023}
\kierunek{Computer Science}
\dziedzina{
11.3 Computer Science\\
}
\klasyfikacja{
  D. Software\\
  D.2. Software Engineering\\
  D.2.4. Software/Program Verification
  }

\title{Accelerating package expansion in Rust through development of a~semantic versioning tool}
\titlepl{Przyspieszenie rozwoju pakietów w języku Rust poprzez rozbudowę narzędzia do
	semantycznego wersjonowania}

\begin{document}
\maketitle

\begin{abstract}
In many programming languages there exist countless nuances, making developers accidentally release
new versions of their packages that are not backwards-compatible. Such releases can directly impact
projects which are using their packages, causing bugs or even compilation errors when using the
latest version. One of the affected languages is Rust, which also lacks (itself) a~built-in
mechanism for enforcing semantic versioning.

The aim of this thesis is to describe the development of a~tool for Rust programmers to reduce the
chances of publishing a~new version of the code that violates semantic versioning.

There are already on-going plans to bundle this tool into the language's standard
development toolchain. It would make it commonly used and therefore help users to safely get
bug fixes, security patches and new functionality, without worrying about their app being broken
by a~dependency change.
\end{abstract}

\tableofcontents


\chapter*{Executive summary}
\addcontentsline{toc}{chapter}{Executive summary}

When new releases of Rust libraries are published, it may happen that unintended public API changes
are made and cause their users' code to suddenly stop compiling. In such cases, the packages'
maintainers have to remove the faulty releases or create patches to fix them and the users
have to temporarily downgrade the library's version. We have discovered that more than 1~in~6
of the most commonly used Rust libraries had this issue. This is a~serious concern as the problem
is alarmingly common but has never been satisfactorily resolved so far.

The following document describes the development of \texttt{cargo-semver-checks}, an~open-source
tool that solves this problem. It detects changes that are
not backwards-compatible in Rust library's code before a~new version is released by its maintainers.

Instead of directly looking at the source code, the tool compares the automatically generated
documentation of two versions of the library's public API. To parse this data, it uses
a~query engine called \texttt{Trustfall} that is capable of reading from any organized database,
allowing for the checks run by the tool to be written in form of short queries. Thanks to
this approach, the project is much more sustainable than other attempts of this kind,
which are not maintained anymore.

We worked in a~team of four as part of our bachelor's thesis on improving the tool to make it
convenient and accessible for the Rust community. We extended the command-line interface
according to the users' requests and needs, developed a~brand new version of
the tool's GitHub Action and implemented various new lints to prevent even more breaking changes.
Additionally we greatly expanded the test suite, fixed multiple bugs that previously
made it difficult for the community to use the tool and further improved the codebase to attract
new contributors.

Furthermore we present the results of running \texttt{cargo-semver-checks} on a~large number
of popular crates. This research proved that the problem is far more common than expected.
Moreover, gathering statistics on occurrences of different semantic versioning breaks
revealed the most frequent ones which the developers should pay closer attention to
in their projects.

The tool has shown a~constant increase in its usage rates, growing up to its
well-established position in the Rust ecosystem. The tool is actively used by some of
the largest Rust libraries such as tokio, libp2p, data-encoding and pest (which have
over 150 million downloads). It also has earned itself the trust of numerous worldwide
known companies including Amazon, Microsoft, Adobe and Mozilla. Their developers
use \texttt{cargo-semver-checks} on a~daily basis to prevent semantic versioning breaks in their
publicly available libraries.

The tool is to become part of the official package manager and build system for the Rust language.
This further reassures us that its future is bright and makes us proud to see it making
such an~impact on the entire world of Rust.


\chapter*{Introduction}
\addcontentsline{toc}{chapter}{Introduction}

Rust is a~relatively new language that has lately been gaining on popularity. Amongst its standard
development toolchain there is a~software called cargo -- Rust's build system and package manager,
which handles building a~crate -- a~package with Rust code. One of cargo's responsibilities is to
search for the newest crate versions, download them and build dependencies (that is, other
libraries / crates that the code uses).

Cargo assumes that all crates follow semantic versioning -- a~set of rules that decide when it is
safe to use the new version of a~library without any problems (the projects that depend on this
library must still compile). Thanks to that, developers of libraries can release patches and
bugfixes and cargo by default will automatically use those newest releases. The exact definition of
semantic versioning is explained in chapter \ref{r:chapter_definitions}.

It turns out that it is not obvious which changes are backwards-compatible and which are not.
There are numerous examples (listed in section \ref{r:section_rust_semver_problems})
where experienced Rust developers made a~mistake (due to insufficient knowledge about the language
or by overlooking something) that changed the public API (Application Programming Interface)
of a~library in such a~way that users' code stopped compiling.

The authors of this thesis present their work on a~tool that detects and notifies about a~subset of
problems that make Rust libraries' public API non-backwards compatible, with the goal to make
problems with semantic versioning less frequent in the Rust community. The exact goals of this
project are described in chapter \ref{r:chapter_vision}.

The project is open-source -- all of the authors' work is visible in public repositories (listed in
\ref{r:section_project_structure}). The project existed before their contributions, thus chapter
\ref{r:chapter_implementation} mainly describes how they have extended the project's
functionalities and fixed its issues. Chapter \ref{r:chapter_conclusion} is a~summary of their
results and the project's impact on the Rust community.


\chapter{Definitions}\label{r:chapter_definitions}

\section{Rust language}\label{r:section_rust_language}

With its first stable release in 2015, Rust is a~relatively new programming language that
puts a~strong emphasis on type-safety, memory-safety, performance and concurrency. Unlike
other memory-safe languages, Rust does not use reference counting nor a~garbage collector.
Instead it tracks the object lifetime of all the program's references during compilation,
using its built-in mechanism named \say{borrow checker}.

Among many benefits and advantages of using Rust, there are three dominant features. Firstly,
it provides the user with a~great performance, both time- and memory-wise, allowing for
writing efficient applications. Secondly, it is reliable -- with extensive type system and
focus on runtime safety, Rust prevents the user from making multiple bugs
already during compilation. Finally, it boosts productivity -- its documentation is
comprehensive and the compiler's error messages are clear and helpful. Rust also provides
an~easy to use, integrated package manager and a~dedicated build tool (named cargo),
an~auto-formatter and many more.

According to surveys \cite{survey} conducted amongst developers, Rust is considered to be the most
enjoyable programming language for the 7th year in a~row (as of 2022 summary, meaning it was always
on top since its release in 2015). Over 87\% of respondents who have previously used Rust would
like to continue programming in it.

Because of the high level of safety it provides, Rust has also become the secondary
language for writing the Linux kernel (with the primary being C). While there have been
attempts to add other languages (e.g. C++ in 1997), Rust is the first one to successfully
make it since the kernel's first release in 1991.

\section{Rust's package manager and build system -- cargo}\label{r:section_cargo}

Cargo is the official package manager for Rust. It is responsible for compiling the user's
packages, downloading their dependencies and uploading them to the selected package registry.
It is important to note that in Rust, packages are often referred to as \texttt{crates} with both
names being equivalent and commonly used. Most of the currently publicly available crates are
stored in the Rust community's default crate registry, \texttt{crates.io}. The role of the
registries is to allow users to easily find appropriate crates for their projects, as well as
publish their own work to be used by others. Each registry contains an~index, which itself has
a~searchable list of all the crates available within its registry. Cargo uses the selected registry
and index to download crates and update dependencies.

One can think of cargo as the Rust analogue of Python's pip (the recommended package manager for
Python). But despite their similar roles,
cargo has many advantages making it a~lot safer than pip:
\begin{itemize}
	\item Cargo's crates are always installed using virtual environments, thus they cannot be
		installed in a~location that would interfere with unrelated code.
	\item When a~user adds a~crate to their project, they can only use its direct,
		public interface.
		There is no possibility to access the interfaces of the dependencies that the added crate
		is using itself.
	\item The dependency's name must also be included in a~special file named \texttt{Cargo.toml},
		not just in the source code file.
	\item To make the package's public interface explicit, Rust provides a~series of tools for
		privacy management. Because of this, it is much more difficult for the user to make their
		project depend on something that the package's author did not intend to offer and could
		make private in future versions, causing a~difficult to detect break on the user's end.
	\item Compatibility is taken seriously in Rust, thus packages' APIs (Applications Programming
		Interfaces) are expected to be backwards-compatible (and to follow semantic versioning,
		described in section \ref{r:section_semver}) when a~new version is released. When a~package
		does not satisfy this requirement, there exists a~procedure called \textit{yanking}.
		It allows to remove the faulty version from the index without deleting any data, but moves
		the index back to the most recent working version. Yet in case an~incompatible package is
		downloaded before being yanked, projects using the package still mostly break during
		compilation or build time rather than runtime.
	\item The packages are also written in Rust and never require anything to work other than the
		package code itself. This minimizes the threat of missing pre-installed
		system dependencies.
	\item Cargo allows for an~executable to contain different versions of the same package,
		enabling the user to include multiple dependencies based on a~shared sub-dependency.
		When multiple versions of the sub-dependency are required, they are all included
		without conflicts.
\end{itemize}

Each package in Rust contains a~special \texttt{Cargo.toml} file, called \textit{manifest}. It is
written in the TOML format for simplicity and contains all the metadata necessary for the package
to compile. Every manifest has numerous sections, with the most important two being:
\begin{itemize}
	\item \texttt{[package]} -- contains the information for cargo to compile the package.
		The minimum that must be provided are the package's name and version. Additional fields
		may be required when the maintainer wants to share their work with others by publishing
		to a~registry.
	\item \texttt{[dependencies]} -- lists all of the package's dependencies. Before compiling,
		cargo has to search for and download all of them. The default search location is crates.io.
\end{itemize}
Presence of the manifest file allows the crates.io registry to display useful information about
a~particular package. The users are provided with a~\texttt{README.md} file, instructions on how
to add the package to their own Cargo.toml file, a~list of all available versions, statistics
regarding the downloads for each version and more.

There also exists a~second file for cargo -- \texttt{Cargo.lock}. It contains detailed
information about the used dependencies. Since all dependencies come from some version control
system, the user may not need to provide an~exact revision of a~dependency in their
\texttt{Cargo.toml} file. In this case, there are two scenarios for what happens with the
\texttt{Cargo.lock} file during compilation:
\begin{itemize}
	\item \texttt{Cargo.lock} is \textit{not} present -- cargo then creates it and for each
		dependency in the manifest downloads the most recent revision available, saving additional
		data about it in the file.
	\item \texttt{Cargo.lock} is present -- for the dependencies already present in the file, cargo
		does not look for their newest version and instead downloads the one specified in the file.
		For dependencies that are not present, it does the same as above except it adds them to
		the file rather than overwrites it.
\end{itemize}
There is a~dedicated command for updating the dependencies -- \texttt{cargo update}.
Upon execution, the \texttt{Cargo.lock} file is either created or updated with the latest
revisions of used dependencies.

Besides downloading dependencies, users can also publish their own packages using
the \texttt{cargo publish} command. To use the command, the user must first authenticate with
an~API token. After doing so, the command first performs preliminary checks, including searching
the manifest for a~key to determine whether the user is allowed to upload their package
to the selected registry (by default crates.io). Once the checks are passed, the package is
uploaded and available publicly.

\section{Semantic versioning (semver)}\label{r:section_semver}

Semantic versioning is a~set of rules and requirements dictating how version numbers for packages
are assigned and incremented. They are based on, but not limited to currently widely used
practices in both open-source and closed software development. For these rules to work, one has to
first declare a~clear and precise public API, changes of which refer to specific incrementations of
the package version number. The number is most often denoted as X.Y.Z (Major.Minor.Patch), where:
\begin{itemize}
	\item \textbf{Major} number is incremented when backwards-incompatible API changes were made,
	\item \textbf{Minor} number is incremented when a~new functionality was added in
		a~backwards-compatible manner,
	\item \textbf{Patch} number is incremented when backwards-compatible bugfixes were made.
\end{itemize}
With these rules, version numbers and their changes reflect actions in the underlying
package's code. By definition, two versions are considered compatible if their leftmost non-zero
component remains unchanged.

The main purpose of semantic versioning is to track the changes of a~package's API and reflect
their severity with appropriate numbering.
When systems with multiple dependencies are considered, the lack of semver can have serious
consequences. If the dependency rules are too tight, one is in danger of facing a~version lock
(inability to release a~new version of one package without new versions of all packages depending
on it). On the other hand, if the rules are too loose, there is a~risk of version promiscuity
(assuring that currently released version will be compatible with too many future releases).
Without semver, such issues can occur and make it difficult for developers to further expand
their projects.

In Rust, semver is used by cargo for specifying package version numbers. This makes for a~common
compatibility convention between different versions of the same package. Cargo assumes that
it is safe to update a~dependency within a~compatibility range without it breaking the build.
This range can be defined in the manifest using the \textit{version requirement syntax}, allowing
the user to select the upper bound for maximum compatible number anywhere from one, specific
version (not allowing any updates), to no limit at all (equivalent to allowing even major updates).
The default behavior is to update until the next major version.

There also exists a~procedure for releases that are not backwards-compatible despite their version
number satisfying it, called \textit{yanking}. The yanked release of a~package is not
deleted as it may be in use by some projects, but forces a~version increment against its own number
whenever the next version is released. The resolver ignores yanked package versions unless they are
already present in the \texttt{Cargo.lock} file. While there are many reasons behind packages being
yanked, one of the main causes remains breaking semver.

\section{Abstract syntax tree (AST)}\label{r:section_ast}

An abstract syntax tree represents the abstract syntactic structure of the text (often source code)
in form of a~tree, where every node denotes an~occurrence of some construct in it. The tree is
called \say{abstract}, because it omits some syntax details, such as parentheses or if statements,
and focuses on structural or contextual details.


\chapter{State of the art}\label{r:chapter_state_of_the_art}

\section{Problems with using semver in Rust}\label{r:section_rust_semver_problems}

It might seem easy to maintain semver, but some violations are hard to notice
when not actively searched for. Consider the following example:
\vspace{-3pt}
\begin{verbatim}
  struct Foo {
      x: String
  }

  pub struct Bar {
      y: Foo
  }
\end{verbatim}
\vspace{-5pt}

Changing {\ttfamily Foo.x} type from {\ttfamily String} to {\ttfamily Rc<str>}
causes semver break, even though it is a~non-public field of a~non-public struct.
That is because {\ttfamily String} implements {\ttfamily Send} and {\ttfamily Sync} traits
that are automatically derived, making both {\ttfamily Foo} and {\ttfamily Bar}
implement {\ttfamily Send} and {\ttfamily Sync}.
In contrary, {\ttfamily Rc<str>} implements neither of them,
so the change results in a~publicly visible struct {\ttfamily Bar} losing a~trait.

The given example is not only unobvious, but also even harder to notice in large codebases, where
those structs could be in completely different locations. In fact, a~similar error crept into the
release v3.2.0 of a~well-known crate maintained by the Rust team -- {\ttfamily clap}. More details
about it can be found in section \ref{r:section_real_life_semver_breaks}.

The same issue almost happened (but has been prevented thanks to our tool) in another common
library \texttt{rust-libp2p}, where it is clear from the conversation \cite{issue-libp2p} that
the maintainers were not expecting their type to stop being \texttt{UnwindSafe} and were likely
not even aware that their type was publicly \texttt{UnwindSafe} to start with.

\section{Consequences of breaking semver}\label{r:section_semver_breaking_consequences}

When a~maintainer publishes a~new version of their crate that is breaking semver,
it is causing a~major inconvenience for the crate's users.
Their code might just stop compiling when the offending version gets downloaded.
This could also happen if the crate containing the violation is not an~immediate dependency,
so one semver break could result in tons of other broken crates.

Debugging a~cryptic compilation error that starts showing up one day,
without any change to the code, can be frustrating. In fact, we have experienced it during our
contributions (one of the tool's users opened a~GitHub Issue \cite{issue-compiling-fails}), as one
of our dependencies broke semver. This is a~major problem, as it might drive the users to stop
using such crate.

Because of that, maintainers have to yank the incorrect releases as soon as possible -- otherwise
more users would encounter this problem and their trust in this particular crate (and crates using
it as a~dependency) would decrease. Even though yanking the release seems easy, fixing the semver
break could also result in a~lot of additional work for the maintainers -- they have to investigate
the semver break when it is reported, inform the users about the yanking and possibly help some
move away from the faulty release.

\section{Real-life examples of semver breaks} \label{r:section_real_life_semver_breaks}

Some of popular Rust crates with millions of downloads happened to break semver:
\begin{itemize}
    \item {\ttfamily pyo3 v0.5.1} accidentally changed a~function signature \cite{pyo3-issue},
    \item {\ttfamily clap v3.2.0} accidentally had a~type stop implementing an~auto-trait
		\cite{clap-issue},
    \item multiple {\ttfamily block-buffer} versions accidentally broke their MSRV contract
		\cite{block-buffer-issue},
    \item and many more. We have developed a~script that scans all releases for semver breaks we
		can detect. The results are covered in section \ref{r:section_results_statistics}.
\end{itemize}

Those were examples of popular crates with experienced maintainers, but the problem is even more
prominent in less used crates where developers might not know the common semver pitfalls. A paper
\cite{paper} claims that out of the yanked (un-published) releases, semver break was the leading
reason for yanking, with a~shocking 43\% rate. It also mentions that 3.7\% of all releases
(and there is more than 300 000 of them already) are yanked, which shows the scale of the problem
-- thousands of detected semver breaks.

\section{Existing tools for detecting semver breaks}\label{r:section_existing_semver_tools}

There are not many great tools for semver checking in existence. The main reason for that is that
the semantics of popular languages make complete and automatic verification practically impossible.
There are some initiatives to combat this. For example, the Elm language \cite{elm-lang} by design
enforces semantic versioning. Its type system enables automatic detection of all API changes.
Outside of that, it does not appear that tools for checking semver in established languages like
Python or C++ are commonly used in the industry.

Unfortunately, the Rust language's semantics were also not designed with semver in mind.
Despite this, there are some existing tools for semver checking. First of them,
\texttt{cargo-breaking}, works on the abstract syntax tree. Although ASTs contain all the
information needed for comparing API changes, it has a~major drawback -- two trees must be
navigated at once. It can get complex and tedious (especially when checking for moved or removed
items), because the abstract syntax tree could change quite a~lot, even without any public API
changes. Another issue is that both language syntax and the structure of the abstract syntax tree
often change along with the development of the language, which makes maintenance time-consuming.

The second existing tool is \texttt{rust-semverver}, which focuses on the metadata present in the
rust-specific rlib binary static library format. Because of that, the user experience is far from
ideal, as it forces the user to use some specific unstable versions of the language, and the
quality of error messages is limited.

In comparison, the cargo-semver-checks' approach to write lints as queries seems to work
really well. Adding new queries is designed to be accessible and the maintenance comes down to
keeping up with rustdoc API changes, which seems to be about as low effort as it could be.


\chapter{Vision}\label{r:chapter_vision}

\section{Project purpose}\label{r:section_project_purpose}

We have taken part in the development of \texttt{cargo-semver-checks} -- a~tool meant to detect
semver breaks of a~Rust project before publishing its new version to the registry. The main goal
of the project is to reduce the number of issues with new releases of libraries, which would deepen
the Rust community's trust in the packages they are using and in the whole ecosystem.

The tool is \textit{not} meant to detect all possible semver breaks. Instead, the detection
mechanism and existing lints are written with a~strong focus on finding only true-positives.
Each false-positive (which is defined as a~scenario where the tool wrongly reports a~semver break)
is perceived as a~bug in the tool. There are two advantages with this approach:
\begin{itemize}
	\item the tool is suited to be used in continuous integration for finding semver issues in
		commits or Pull Requests -- it reports semver breaks only when its output should not be
		ignored and requires manual inspection,
	\item we are avoiding a~scenario in which a~user encounters false-positives, becomes
		frustrated and their trust in the tool is reduced, making them less likely to use the
		tool again.
\end{itemize}

Similarly to the Rust compiler, the reports contain detailed information about the issues,
including the line of code in which each one was found, both short and long description of the
triggered lint and, if possible, reference links to the Rust documentation describing the semver
issue in more detail.

\section{Plans to merge the tool into cargo}\label{r:section_merge_into_cargo_plans}

There are already on-going plans to merge cargo-semver-checks into the official Rust development
toolchain, which has been initiated by one of the developers working on cargo
\cite{issue-merge-cargo}. That would make the tool available to all Rust programmers through
a~cargo subcommand and would be a~huge step towards reducing the number of package releases with
semver issues.

Because of those plans, the command-line interface of the project is similar to the interface of
cargo's subcommands. Additionally, cargo developers take part in discussions about the design of
the interface \cite{issue-cli-interface} or sometimes even review changes to the code or develop
new functionalities.

cargo-semver-checks was the one tool chosen to be merged into cargo, mainly because it is the
easiest to maintain (as described in section \ref{r:section_existing_semver_tools}) and does not
report false-positives.

\section{Project usage}\label{r:section_project_usage}

The basic functionality of the tool is to compare two versions of the code (the \say{current} crate
and the \say{baseline} crate) and notify the user about the semver issues that were found when
checking the two versions.

One can run the tool by directly passing the path to the current crate and either getting the
baseline crate from registry, or by directly passing a~path to the baseline crate.

Because a~large portion of libraries are currently developed using GitHub, we are also providing
a~continuous integration job via GitHub Actions to automatically ensure that a~Pull Request
satisfies semver.

\subsection{Running locally}\label{r:subsection_running_locally}

The tool can be used on a~local copy of the project (e.g. just before releasing its new
version to the registry through \texttt{cargo publish}) by passing the path to its manifest.
The baseline crate can be passed in a~similar manner, but alternatively it is possible to either
specify a~version of the crate in the registry with which the current crate should be compared
with, or (assuming that the current directory is a~git repository) by passing a~git revision where
the baseline project is located.

Additionally, it is possible to check all crates in a~workspace (which is a~collection of crates)
one by one by passing the path to the manifest that defines the workspace.

\subsection{Usage in continuous integration}\label{r:subsection_usage_in_ci}

Seeing that GitHub is the most popular internet hosting service for version control amongst the
Rust libraries developers, one of the goals of the project is to implement a~GitHub Action that
checks whether a~given git branch has not violated semver with recent changes.

There are two reasons as to why this continuous integration job is beneficial for the developers:
\begin{itemize}
	\item it can be used together with a~job that automatically publishes a~new version of the
		library when it passes the semver lints and the version of the package has been raised
		in the manifest file,
	\item it provides important information for library maintainers for deciding when a~Pull
		Request should be merged -- in case the branch contains a~minor or major change, the
		maintainers could want to wait with merging it until they plan to make a~minor or major
		release of their library.
\end{itemize}

\section{Project baseline}\label{r:section_project_baseline}

Before our contributions, the project was already partially functional, but it had numerous issues
which often prevented the community from adopting the tool into their workflow:
\begin{itemize}
	\item the project did not have many lints,
	\item the code was bugged in multiple ways and lacked some functionality,
	\item the community was not satisfied with the implemented GitHub Action to the point
		where some developers coded their own continuous integration job using
		just the command-line interface of the tool,
	\item some existing lints had false-positives,
	\item the codebase was not in a~state where new contributors could easily begin making changes
		to the project (which is crucial for the project to flourish in the long term).
		For example, adding new lints and tests was not intuitive and required many manual steps,
		the filenames and variable names were not always descriptive enough and the code lacked
		comments that explained some of the logic and decisions behind it.
\end{itemize}


\chapter{Theory}\label{r:chapter_theory}

\section{Project structure}\label{r:section_project_structure}

In order to check two versions of a~crate for semver violations, the tool needs a~consistent method
for generating the data to be compared. This is achieved by using rustdoc (from the official
Rust toolchain) to generate the documentation for Rust projects. However, the documentation itself
often changes its format and thus is not stable nor organized enough to be directly used by
cargo-semver-checks. For this reason, the tool uses Trustfall to access and query the output
received from rustdoc.

Trustfall is a~special query engine, designed to query any data source or a~combination of data
sources such as APIs, raw files (JSON, CSV, etc.), git version control, web pages and many more.
Its main idea is to represent the data as an~organized graph with vertices and edges that the
engine can traverse. To navigate during the search, it follows consecutive filtering rules defined
by the user. In cargo-semver-checks, these rules are contained within \texttt{.ron} files which
are called \texttt{lints} inside the project (more detail in section
\ref{r:section_cargo_semver_checks}).

Despite being the project's core component, new lints are not the only thing we developed. In total
there are four repositories that we have been actively contributing to for the last few months:
\begin{itemize}
	\item \textbf{cargo-semver-checks} -- a~CLI (command-line interface) tool that runs the lints
		(Trustfall queries),
	\item \textbf{trustfall-rustdoc-adapter} -- implements the Trustfall interface for one specific
		rustdoc version,
	\item \textbf{trustfall-rustdoc} -- allows running Trustfall queries over rustdoc regardless of
		the currently used rustdoc version,
	\item \textbf{cargo-semver-checks-action} -- the tool in form of a~GitHub Action that can
	    be added to GitHub repositories.
\end{itemize}

\begin{figure}[h]
	\centering
	\includegraphics[width=\linewidth]{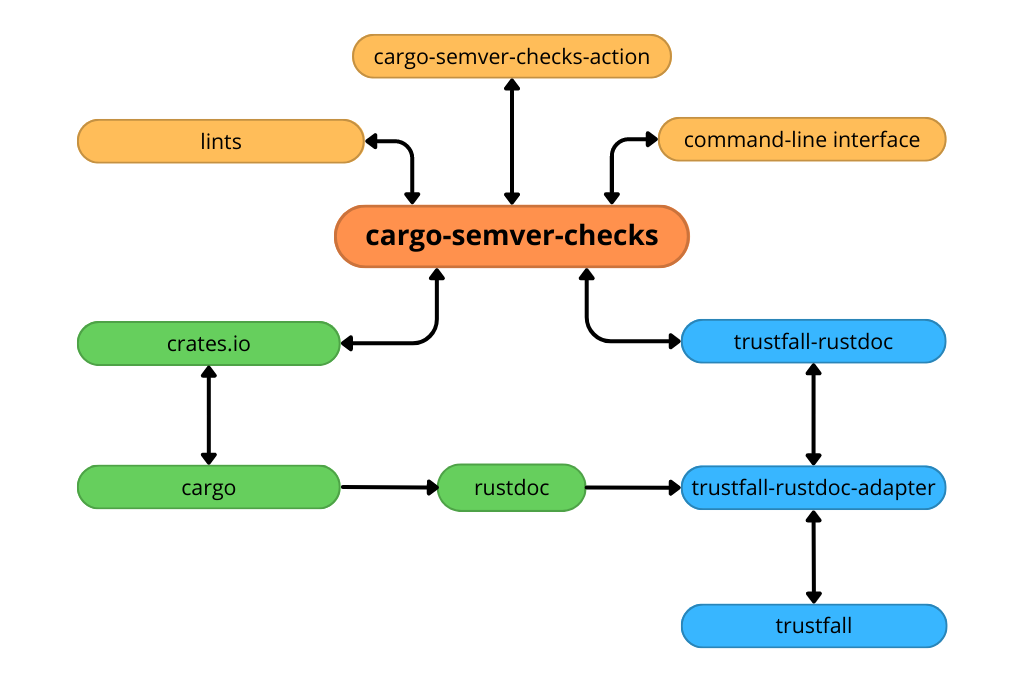}
	\caption{Diagram of the project structure.}
\end{figure}

To learn more about the project's individual components present in the diagram, see:
\begin{itemize}
	\item cargo-semver-checks -- chapters \ref{r:chapter_vision}
		and \ref{r:chapter_implementation}, section \ref{r:section_cargo_semver_checks}
	\item lints -- sections \ref{r:section_cargo_semver_checks} and \ref{r:section_new_lints},
	\item command-line interface -- section \ref{r:section_cli},
	\item cargo-semver-checks-action -- sections \ref{r:section_cargo_semver_checks_action}
		and \ref{r:section_github_action},
	\item cargo -- section \ref{r:section_cargo},
	\item crates.io -- section \ref{r:section_cargo},
	\item rustdoc -- section \ref{r:section_project_structure},
	\item trustfall -- section \ref{r:section_project_structure},
	\item trustfall-rustdoc-adapter -- section \ref{r:section_trustfall_rustdoc_adapter},
	\item trustfall-rustdoc -- section \ref{r:section_trustfall_rustdoc}.
\end{itemize}

\section{cargo-semver-checks}\label{r:section_cargo_semver_checks}

The goal of cargo-semver-checks is to detect semver violations within Rust crates. Given two
versions of the same crate:
\begin{itemize}
	\item \textbf{baseline} -- the existing version that defines a~public API,
	\item \textbf{current} -- the modified version of the baseline that is vulnerable to
		breaking semver,
\end{itemize}
the tool uses Trustfall as a~layer of abstraction that remains constant regardless of changes in
the underlying JSON representation format (functions, structs, enums, variants, etc.) To better
visualize this concept, one can imagine a~query that asks Trustfall to \say{Find all public Enums
from the baseline version that no longer exist in the current version of the crate}. That kind
of query never makes any statements or assumptions about its data representation or format.

After that, the versions are compared using all of the implemented lints to find the breaking
public API changes. They are the core component of the project and are carefully designed so that
one break can only be reported by one, precise lint. This is especially important because we want
the user to be provided with an~extensive, coherent and clear information about the error they
have made.

In cargo-semver-checks, each lint contains a~single \texttt{SemverQuery} to execute, consisting of
the following fields:

\begin{itemize}
	\item \texttt{id} -- a~short, unique name identifying the lint, most often matching the
		filename that contains the lint,
	\item \texttt{human\_readable\_name} -- a~longer, human-readable name,
	\item \texttt{description} -- general information about what is the detected problem,
	\item \texttt{required\_update} -- the minimum version bump needed to avoid breaking semver,
	\item \texttt{reference\_link} -- a~link to the official Rust documentation corresponding to
		the issue (when available),
	\item \texttt{query} -- the Trustfall filtering rules responsible for comparing two versions
		and detecting semver violations (described below in more detail),
	\item \texttt{arguments} -- optional, user-defined Trustfall constants for the query,
	\item \texttt{error\_message} -- a~longer and more technical information, telling the user
	    about the exact semver violation that occurred and what could have caused it,
	\item \texttt{per\_result\_error\_template} -- the name of the item that broke semver and
	    its exact position in the code.
\end{itemize}

\begin{figure}[h]
	\centering
	\includegraphics[width=\linewidth]{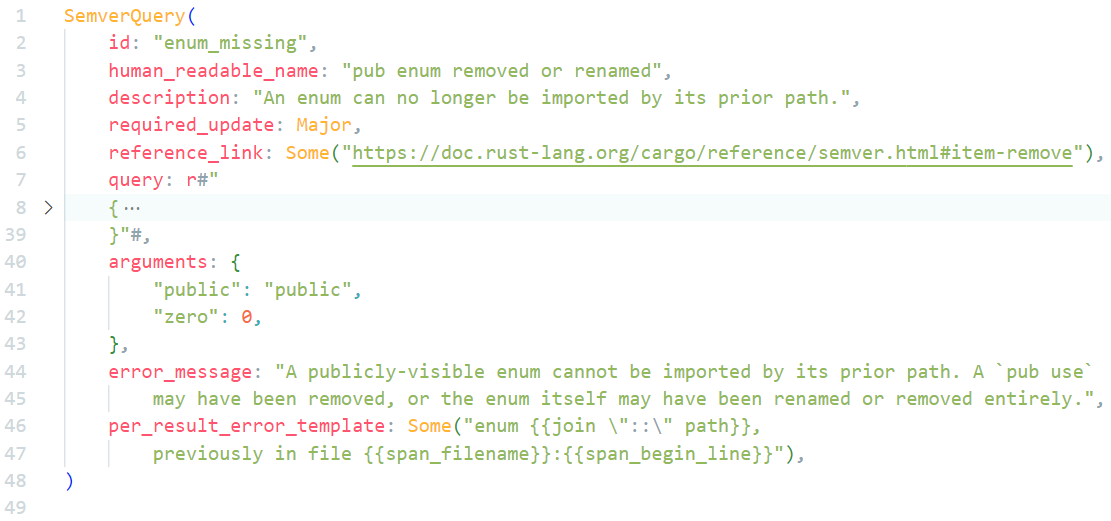}
	\caption{The \texttt{enum\_missing} lint (for its entire query, see figure
		\ref{fig:enum_missing_lint_query})}
	\label{fig:enum_missing_lint}
\end{figure}

Each query defines the filtering rules necessary to compare the baseline and current versions of
a~crate. They contain elements such as:
\begin{itemize}
	\item \texttt{@tag} -- allows the value to be used elsewhere in the query by
		tagging its name,
	\item \texttt{@filter}, \texttt{@fold}, \texttt{@transform} -- some of the available Trustfall
		directives that modify the sets of items processed by the query,
	\item \texttt{@output} -- specifies which values are returned by the query,
	\item \texttt{@optional} -- edges marked optional are allowed to not exist, in which case
		\texttt{@output} properties within become \texttt{null} instead of causing that result
		to be discarded,
	\item \texttt{... on Enum} -- \texttt{...} narrows that location in the query to the specified
		kind of items (in this case Enums), discarding all results that had other types of data
		in that location
	\item \texttt{\$public}, \texttt{\$zero} -- \texttt{\$} refers to the argument with the same
		name,
	\item \texttt{\%path} -- \texttt{\%} refers to a~tagged element with the same name.
\end{itemize}

\begin{figure}[h]
	\centering
	\includegraphics[width=0.975\linewidth]{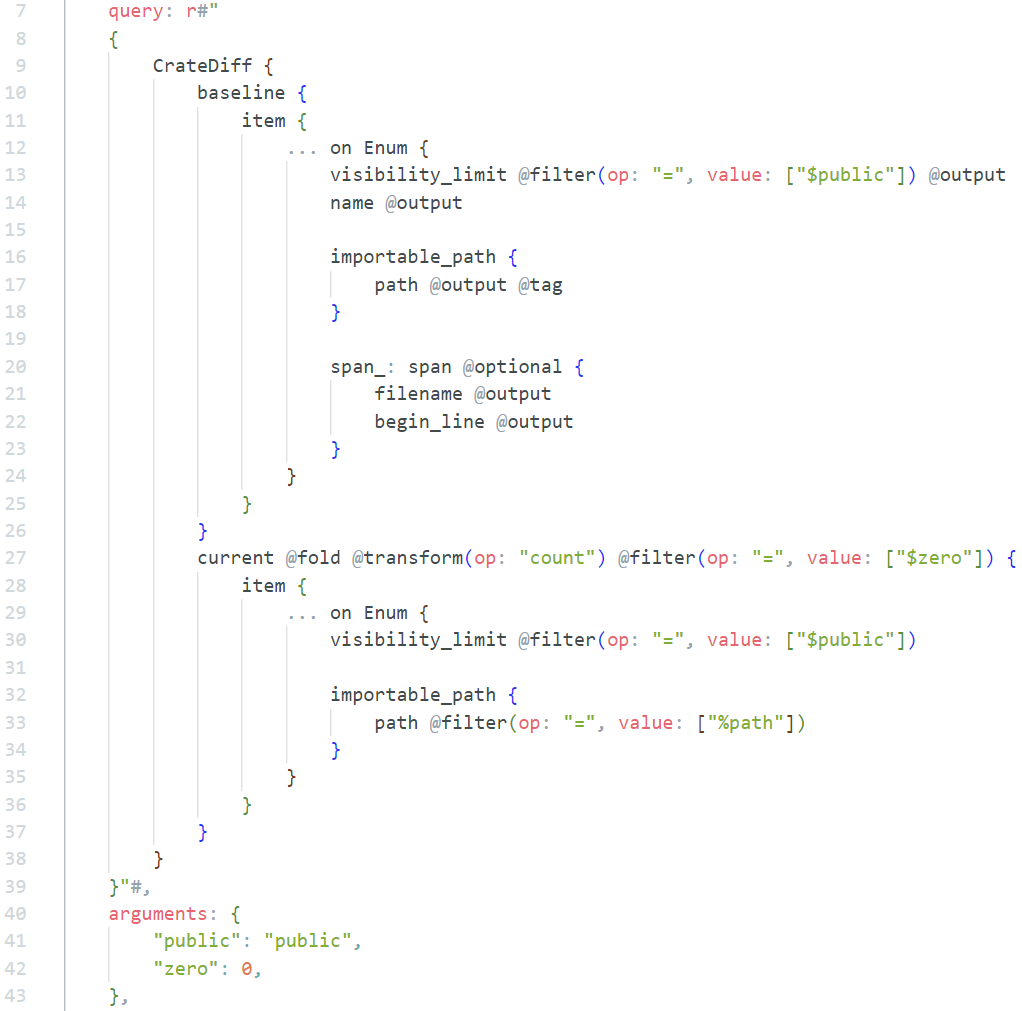}
	\caption{The \texttt{enum\_missing} lint's query}
	\label{fig:enum_missing_lint_query}
\end{figure}

The presented query is part of the lint seen in figure \ref{fig:enum_missing_lint}.
During its execution, Trustfall successively:
\begin{itemize}
	\item collects all of the baseline crate's public Enums, tagging their paths,
	\item collects all of the current crate's public Enums such that their path is present among
		the baseline version's tagged paths,
	\item counts the occurrences of the filtered current Enums among the filtered baseline Enums,
	\item filters the current Enums by their resulting count being zero,
	\item returns the Enums that passed the last filtering.
\end{itemize}
Just like that, the query has returned all of the public Enums that were present in the baseline
version of the crate, but got removed in the current version.

\section{trustfall-rustdoc-adapter}\label{r:section_trustfall_rustdoc_adapter}
The output generated by running rustdoc on a~crate is a~tree and can be in one of the two formats:
\begin{itemize}
	\item HTML -- the default option, formatted with CSS, designed for users to read (however it is
		difficult to parse),
	\item JSON -- convenient for further processing, but practically unreadable to humans due to
		its complex structure and large size of up to a~few hundred MB.
\end{itemize}
This makes the JSON format a~natural candidate for running Trustfall queries. However, it still has
to be transformed into a~structure that is understandable for Trustfall, namely a~graph in which
vertices are assigned types and type-dependent properties.

This is where trustfall-rustdoc-adapter comes in. It provides an~interface (called schema)
for querying the rustdoc's output by associating JSON nodes with Trustfall vertices.
To achieve this, the adapter implements the following methods:
\begin{itemize}
	\item \texttt{resolve\_starting\_vertices} -- returns an~iterator over the starting vertices of
		the graph (entry points for query executions),
	\item \texttt{resolve\_property} -- resolves the value of a~specified vertex' property,
	\item \texttt{resolve\_neighbors} -- finds the neighboring vertices across an~edge,
	\item \texttt{resolve\_coercion} -- attempts to coerce a~vertex into a~more specific type.
\end{itemize}
These are used by the Trustfall engine to execute lint queries written in the Trustfall language.

\section{trustfall-rustdoc}\label{r:section_trustfall_rustdoc}

Rustdoc output format is under constant development and changes between Rust versions.
Since the tool is supposed to be widely used, it is important to ensure that it works with
the recent versions of Rust (and consequently with all recent versions of rustdoc).
Therefore, the trustfall-rustdoc-adapter is developed on several different git branches,
each of them handling a~specific rustdoc version.

The trustfall-rustdoc brings them together in the form of a~single tool. Given the path to
the rustdoc JSON output, trustfall-rustdoc loads the file, detects which rustdoc version was used
and chooses the corresponding trustfall-rustdoc-adapter branch to run the Trustfall queries.
This provides a~version-independent interface which is used by cargo-semver-checks.

\section{cargo-semver-checks-action}\label{r:section_cargo_semver_checks_action}

Using cargo-semver-checks in a~continuous integration workflow consists of three main steps:
setting up the Rust toolchain, installing cargo-semver-checks, and finally -- running it
on a~crate. The aim of the cargo-semver-checks-action is to simplify this process and make it
a~single step, hiding the technical details from the user.

The cargo-semver-checks-action is implemented as a~GitHub Action, which is a~way of creating
a~reusable GitHub workflow. It is hosted on a~public repository and can be used in continuous
integration by referencing its name and version in the workflow file. Besides being run
with default parameters, the action can be configured by providing optional inputs (such as
the name of the crate to check), which are passed to the CLI of the cargo-semver-checks binary.

The V1 version of the action was implemented as a~bash script, which was a~quick, but difficult
to maintain solution, which allowed the action to be run only on Linux hosts. This is why
we implemented the V2 version of the action in a~platform-independent way using TypeScript
and Node.js, allowing it to be used on all runners supported by GitHub Actions.
The differences between the V1 and V2 of the action are further discussed in section
\ref{r:section_github_action}.


\chapter{Implementation}\label{r:chapter_implementation}

\section{The lint set expansion}\label{r:section_new_lints}

The continuously expanding list of lints is the heart of the project.
Before our contributions, there were 18 existing lints.
We have managed to implement many new ones:
\begin{itemize}
    \item {\ttfamily enum\_marked\_non\_exhaustive}
        -- detects marking an~{\ttfamily Enum} as {\ttfamily \#[non\_exhaustive]},
		removing the ability to match on them;
    \item {\ttfamily enum\_struct\_variant\_field\_missing}
        -- detects missing fields in a~struct variant of an~{\ttfamily Enum},
        which changes how the variant must be initialized;
    \item {\ttfamily function\_parameter\_count\_changed}
        -- detects functions with changed parameter count,
        changing the way the function must be called;
    \item {\ttfamily tuple\_struct\_to\_plain\_struct}
        -- detects changing struct kind from tuple to plain,
        changing the way the struct must be initialized;
    \item {\ttfamily trait\_unsafe\_added}
        -- detects marking a~{\ttfamily Trait} as {\ttfamily unsafe},
        removing the ability to do a~safe {\ttfamily impl};
    \item {\ttfamily trait\_unsafe\_removed}
        -- detects removing {\ttfamily unsafe} marker from a~{\ttfamily Trait},
        removing the ability to do an~unsafe {\ttfamily impl};
    \item {\ttfamily enum\_must\_use\_added}
        -- detects marking an~{\ttfamily Enum} with the {\ttfamily \#[must\_use]} attribute,
        removing the ability to ignore the enum when returned from a~function;
    \item {\ttfamily struct\_must\_use\_added}
        -- detects marking a~struct with the {\ttfamily \#[must\_use]} attribute,
        removing the ability to ignore the struct when returned from a~function;
    \item {\ttfamily trait\_must\_use\_added}
        -- detects marking a~{\ttfamily Trait} with the {\ttfamily \#[must\_use]} attribute,
        removing the ability to ignore a~function's result, when it is returning an~object
        with dynamic type that implements this trait;
    \item {\ttfamily function\_must\_use\_added}
        -- detects marking a~function with the {\ttfamily \#[must\_use]} attribute,
        removing the ability to ignore the function's result;
    \item {\ttfamily inherent\_method\_must\_use\_added}
        -- detects marking a~method with the {\ttfamily \#[must\_use]} attribute,
        removing the ability to ignore the method's result.
\end{itemize}
To enable the new lints, we had to do some development of the Trustfall rustdoc adapter.
It mainly resolved around extending its schema and implementing the changes.
The challenge was to design the new interface in a~way that would
still make it possible to extend it without problems in the future.

One of the notable changes in the schema involved attributes.
It originally exported its value as a~string, which made traversing
its inner tags in lints not possible.
The solution was to refactor it into a~proper tree structure.
This meant that we had to introduce a~breaking change there
and do minor adjustments in the existing lints,
making it more future-proof at the same time.

\section{Command-line interface improvements}\label{r:section_cli}

The main change which we implemented was adding the ability to specify
features that a~crate should be built with for the checks. It is also
possible to use different configuration for the current and baseline version.
This is important because some functionalities may change between features. What is more,
crates often have a~feature that breaks semver by design, which
we discovered when working on the semver-crater \ref{r:section_results_semver_crater}.
Now the users can adjust the set of features for their specific use cases.

The new default behavior is to include all the features,
filtering out heuristically the ones that break semver by looking at
the feature's name. For example, a~feature named {\ttfamily unstable} would not be
included. The list of banned names was created by looking
at different features in popular crates.

It was not easy to correctly handle all edge cases,
thus we wrote a~heavy test suite for this change.
For example, one of the optimizations written in cargo-semver-checks
is caching the generated rustdoc and we had to make sure that
the caching mechanism is not used when a~user changes the enabled features
(which might change the rustdoc).

We also improved the user experience by spotting and fixing a~couple bugs
in the command-line interface -- both in the tool's behavior
in combination with specific flags and in its output messages.
For details, see section \ref{r:section_bugfixes}.

\section{Brand new GitHub Action}\label{r:section_github_action}

The V1 version of cargo-semver-checks-action was created when
cargo-semver-checks lacked a~built-in logic of choosing the correct baseline version and was not
generating the rustdoc output automatically. Both of these tasks were therefore handled
by the action. When our work on the project started, the tool had already gained these features
making the action obsolete. Its only advantage over the manual configuration was that it installed
both the Rust toolchain and the tool itself. However, it downloaded the baseline version from
the repository instead of using the latest release published on crates.io and did not handle
the edge cases properly. The other problem was that the action did not allow passing any parameters
to the tool, which was pointed out by the users who often decided to setup the tool manually
for this reason. Finally, the action was implemented as a~bash script, which made it both difficult
to maintain and runnable only on Linux hosts.

The need for a~redesign was therefore clear. After reviewing the available options,
we decided to use the fact that a~GitHub action might
be run using Node.js and implement the V2 version in TypeScript. It is a~standard way
of creating complex actions, as it has multiple advantages:
\begin{itemize}
	\item TypeScript action is platform-independent and can be run on all runners supported
		by GitHub Actions,
	\item using Node.js makes it possible to depend on a~variety of \texttt{npm} packages,
		in particular the GitHub Actions Toolkit \cite{github_actions_toolkit}, instead of building
		the components from scratch,
	\item TypeScript code is more readable and easier to organize than a~bash script
		or a~GitHub workflow file.
\end{itemize}
The new version of the action not only fixes the problems of the previous V1, but also
adds numerous features:
\begin{itemize}
	\item input \texttt{rust-toolchain}, which allows the user to specify the Rust toolchain
		to use instead of the default \texttt{stable},
	\item inputs \texttt{package} and \texttt{exclude}, both accepting lists of package names,
		allowing to define the exact set of packages to be checked,
	\item input \texttt{manifest-path}, which allows to specify the path to the \texttt{Cargo.toml}
		file of the crate or workspaces to be processed if it is not located in the directory where
		the action is run,
	\item input \texttt{verbose}, which might be used to enable extended output of the tool,
		providing details about every executed lint,
	\item input \texttt{release-type}, with possible values \texttt{major}, \texttt{minor}
		and \texttt{patch}, allowing to specify the type of the release instead of deducing it from
		the crate's version change.
\end{itemize}
Moreover, the action now includes several optimizations of the running time:
\begin{itemize}
	\item cargo-semver-checks tool is downloaded as a~pre-compiled binary instead of being built
		from source,
	\item the rustdoc of the baseline version is cached between the action's runs, so that
		it is generated only once. The caching strategy might be adjusted by the user using inputs
		\texttt{shared-key} and \texttt{prefix-key} (e.g. to share the cache between different
		GitHub jobs),
	\item when possible, downloads from crates.io are made using recently developed \say{sparse}
		protocol \cite{crates_io_sparse_protocol}.
\end{itemize}

Along with the new version of the action, we also developed an~extensive suite of action tests
(not to be confused with the test suite of cargo-semver-checks). The current CI workflow
of the project consists of over 30 checks, which are responsible for:
\begin{itemize}
	\item making sure the action's code is properly formatted and builds without errors
		and warnings,
	\item running the unit tests,
	\item verifying whether the action works on Linux, Windows and macOS with different versions of
		Rust installed,
	\item testing all of the action's inputs by running it on a~test repository,
	\item comparing the cache created by the action with the expected one.
\end{itemize}
Such a~comprehensive collection of tests helped us to find and fix several bugs not only
in the action, but also in cargo-semver-checks itself, and will make it easier to maintain
the project in the future.

\section{Better test suite with integration tests}\label{r:section_test_suite}

Before our contributions, each lint had its own, dedicated test case consisting of:
\begin{itemize}
\item the Rust source code that could be compiled in two different ways (producing
\say{current} and \say{baseline} code),
\item a~file with the results of executing its Trustfall query over this source code.
\end{itemize}
We have replaced this system with a~new one that contains numerous \say{crate pairs}
(two separate directories called \say{current} and \say{baseline} with their own
manifests and source code) and Trustfall query results for each lint over all
available crate pairs. This has brought numerous benefits:
\begin{itemize}
	\item testing each lint on all crate pairs found multiple false-positives by running
		on code originally meant to test other lints
		(e.g. we found that a~lint has erroneously claimed that a~trait
		has been deleted, when in fact it was still present but its definition
		was changed, which could cause user confusion and frustration at the
		misleading information),
	\item the manifests of the baseline and current crates can be different,
		which allows to write tests for lints checking the manifests
		(such lints are not yet implemented, but it is one of the future goals
		of the project),
	\item because we replaced the heavy use of conditional compilation with separate crates
		with regular source code, it is now easier to triage issues like false-positives
		and to add new test cases.
\end{itemize}
In the process of making the test suite more thorough,
we discovered a~variety of edge cases in the underlying rustdoc tool,
which is part of the Rust toolchain and is also under active development.
Since our work made our tests quite extensive,
the rustdoc team is working on ways to integrate our test suite into rustdoc's own,
by re-running our tests on new rustdoc versions \cite{issue-rustdoc-our-tests}.

We also wrote a~new kind of test to ensure that some bugs found by us or reported by the users
have been fixed. Such tests were made by writing GitHub workflow jobs (which are triggered by every
commit and Pull Request) that downloads the tool and the user's git repository,
runs the tool just like the user did to detect the bug
and checks that it is no longer present.
Thanks to this, we can be sure that the user who reported the bug
will not experience it ever again.
Just the writing of such tests helped us find other bugs
-- we tried to write tests that run the tool in different ways
with various command-line arguments and they detected that it wrongly handled some arguments.

The integration tests were more appropriate to be written as a~GitHub workflow job
-- cloning the user's git repository takes time and memory, requires internet connection
and often the projects are platform-specific,
thus the tests are not suited to be run locally.
But we also had some tests which were better suited to local runs alongside the crate pair tests
-- they provide feedback immediately and it is easier to debug when such tests fail.
For example, there was a~bug which caused the tool to not check any lints at all
because they accidentally were not being included in the produced binary
(which fortunately was noticed before going to production).
To be sure that this will not happen again, we wrote an~integration test that
locally executes the produced binary and verifies its output.

\section{Healthier codebase}\label{r:section_healthier_codebase}

Our contributions to the tool were not just limited to developing new code. We are aware of
how important it is to maintain high quality of the project's codebase, thus we introduced
several changes and refactors that not only made it easier to navigate and expand, but also
prepared the tool to be even more successful in the future.

With lints being the core of cargo-semver-checks, our first major refactoring targeted both their
general file structure and development process. Imagine a~user who is eager to contribute
and decides to develop a~new lint:
\begin{itemize}
	\item \textbf{Previously} they had to go through a~tedious process of modifying single lines
	of code in several different files. What is more, tests for the new lint had to be written
	in a~single file where all the baseline and current code fragments had to be preceded by
	a~differentiation macro. Although there were instructions available, they lacked
	a~general introduction of the project's approach and design goals.

	\item \textbf{Currently} the user only has to choose an~appropriate name for their lint
	and implement it together with suitable tests -- all the minor file changes are handled
	by a~dedicated script that we developed. The tests have been split into two separate
	directories and allow the user to provide the baseline and current test versions just
	as if they were developing a~crate (with the difference being that instead of avoiding semver
	violations they are supposed to make ones that trigger their lint). The instructions received
	an~overhaul as well to be more welcoming and informative, greatly improving the user experience
	and encouraging them to make new contributions.
\end{itemize}
Shortly after we additionally changed how those new tests are run
(for more details on the tests rework, see section \ref{r:section_test_suite}).

Later on we also refactored the \texttt{baseline.rs} file to make future additions easier.
With this we were able to change how the tool generates rustdoc for the baseline version of a~crate
by introducing a~placeholder project with a~direct dependency to it, imitating how one would
naturally access it.

Besides there were numerous minor changes, such as:
\begin{itemize}
	\item renaming the project's file and function names so that they became more descript,
	\item adjusting the clippy (cargo's built-in formatter) lints,
	\item multiple, smaller refactors and optimizations in various parts of the codebase.
\end{itemize}

\section{Bugfixes}\label{r:section_bugfixes}

Another important part of our work was diagnosing the issues present in the tool.
Throughout our contributions to the project, we managed to fix the following bugs:
\begin{itemize}
	\item current and baseline's rustdoc being generated with different features, which
		resulted in false-positives,
	\item tests failing on Rust 1.67.0 due to the addition of a~public \texttt{rustc}
		internal trait named \texttt{StructuralEq} which is not relevant for the users,
	\item trustfall-rustdoc-adapter incompatibility with the new version of rustdoc v23
		which stopped exporting inlined items of external crates,
	\item rustdoc generation of \say{current} with dependency versions taken from
		\texttt{Cargo.lock} instead of the more intuitive choice of taking them
		from \texttt{Cargo.toml}, which made users confused,
	\item false-positives from target-specific optional features,
	\item compilation error which happened because of semver-incompliant changes
		in one of the tool's dependencies,
	\item the rustdoc of the current and the baseline crates overwriting each other,
	\item CLI choosing the latest release from the registry as a~baseline version
		even when it had been yanked,
	\item CLI ignoring the \texttt{-{}-release-type} option,
	\item CI job wrongly checking the exit code in one of
		the integration tests,
	\item incorrect baseline version printed when the baseline was specified through
		a~path to the manifest,
	\item broken links appearing in the tool's output,
	\item rustdoc command's output showing up without colors.
\end{itemize}
Apart from that, we reported a~few bugs that were or are to be fixed by other contributors:
\begin{itemize}
	\item incorrect path to the generated rustdoc when passing a
		manifest path in the CLI,
	\item the tool failing to work when generating rustdoc of a~crate
		that has only yanked releases,
	\item a~false-positive related to re-exporting types, values and
		macros with the same names,
	\item CLI of v0.18.2 not looking for the baseline in the registry
		as it was supposed to do,
	\item the pre-compiled tool crashing on Windows machines,
	\item output of the tool on the bevy-audio crate seeming
		like it was producing a~false positive,
	\item duplicated lint's output on the bevy-render crate,
	\item the tool printing some internal data while checking the clap
		crate.
\end{itemize}


\chapter{Tool's findings in popular crates}\label{r:chapter_semver_crater}

\section{Statistics on real-world violations of semver}\label{r:section_results_statistics}

Thanks to cargo-semver-checks, we were able to find numerous semver violations
amongst the one thousand most downloaded Rust crates.
This section is a~presentation of all the results we got. To learn more about
our methodology, see section \ref{r:section_generating_witnesses}.

We considered only non-yanked releases of crates which were published
after 01/01/2017 and are not major version updates compared to the prior release. From those
releases, \mbox{919 did not} compile on our machines with the current version of Rust.
Thus, we considered \mbox{14 389} releases in total (which is also the number of times
cargo-semver-checks has been run).

From those non-yanked, non-major releases which we were able to compile, we found that
on average \textbf{every 1 in 30 releases had at least one semver violation}
-- we found one or more violations in 470 of them.
All in all, we found semver violations in slightly more than \mbox{\textbf{1 in 6 crates}}
-- that is, in 173 crates of the 1000 crates we scanned.

In total, we discovered 3305 different instances of semver violations across all releases.
We found that the same release frequently contained more than one instance of the same kind
of semver violation -- e.g. a~release could have removed multiple functions from the public API.
Ignoring multiple instances of the same semver violation in the same release, the semver lints
were triggered 655 times across all releases.

We verified each reported semver violation by constructing a~\texttt{witness} -- a~short Rust code
proving that a~particular non-major release indeed violates semver.
Each witness can be compiled using the earlier release of a~crate,
but results in a~compilation error for the next crate's release.
Moreover, for each result we have examined its source code to confirm
that it is part of the public API accessible by the user and not marked \texttt{\#[doc(hidden)]}.
Thus, we are confident that each result is a~true-positive report pointing out a~real
semver violation.

Interestingly, some changes that are theoretically breaking in the Rust language are commonly
accepted by the majority of the community. The tool seeks to be up-to-date with the current
conventions, but as they change over time, running cargo-semver-checks on older releases can
occasionally produce unintuitive results. For example, the developers used close imitations of the
\texttt{\#[non\_exhaustive]} attribute prior to its official introduction. But in order to use the
newly added attribute instead, they had to purposely make breaking changes in their crates.
However, this is the case with only a~minority of semver violations. The general difficulty of
semver remains unchanged regardless of the ever-changing Rust versions and conventions. This makes
automated tools like cargo-semver-checks an~even more valuable assistance for the developers.

The following table shows for each lint how often it was triggered:

\begin{center}
	\begin{longtable}{| p{7.75cm} | p{1.975cm} | p{1.975cm} | p{1.975cm} |}
		\hline
			&
			\multicolumn{3}{c|}{no.\:of detected violations, counted per} \\
		\cline{2-4}
			\multicolumn{1}{|c|}{\textbf{lint}} &
			\multicolumn{1}{c|}{\textbf{individual}} &
			\multicolumn{1}{c|}{\textbf{different}} &
			\multicolumn{1}{c|}{\textbf{affected}} \Tspace \\
			&
			\multicolumn{1}{c|}{\textbf{items}} &
			\multicolumn{1}{c|}{\textbf{releases}} &
			\multicolumn{1}{c|}{\textbf{crates}} \\
		\hline
			\texttt{inherent\_method\_missing} & $791$ & $41$ & $27$ \Tspace \\
			\texttt{enum\_variant\_added} & $387$ & $140$ & $60$ \\
			\texttt{constructible\_struct\_adds\_field} & $343$ & $123$ & $34$ \\
			\texttt{auto\_trait\_impl\_removed} & $318$ & $57$ & $45$ \\
			\texttt{struct\_missing} & $291$ & $66$ & $40$ \\
			\texttt{function\_missing} & $265$ & $50$ & $33$ \\
			\texttt{inherent\_method\_must\_use\_added} & $239$ & $4$ & $4$ \\
			\texttt{inherent\_method\_const\_removed} & $139$ & $5$ & $3$ \\
			\texttt{derive\_trait\_impl\_removed} & $115$ & $11$ & $11$ \\
			\texttt{enum\_variant\_missing} & $112$ & $27$ & $18$ \\
			\texttt{struct\_pub\_field\_missing} & $79$ & $32$ & $16$ \\
			\texttt{enum\_missing} & $78$ & $26$ & $20$ \\
			\texttt{trait\_missing} & $45$ & $24$ & $19$ \\
			\texttt{method\_parameter\_count\_changed} & $22$ & $14$ & $12$ \\
			\texttt{enum\_marked\_non\_exhaustive} & $17$ & $4$ & $4$ \\
			\texttt{struct\_repr\_c\_removed} & $12$ & $3$ & $3$ \\
			\texttt{constructible\_struct\_adds\_private\_field} & $9$ & $7$ & $6$ \\
			\texttt{inherent\_method\_unsafe\_added} & $9$ & $3$ & $3$ \\
			\texttt{function\_parameter\_count\_changed} & $8$ & $4$ & $4$ \\
			\texttt{function\_unsafe\_added} & $8$ & $2$ & $2$ \\
			\texttt{unit\_struct\_changed\_kind} & $5$ & $3$ & $2$ \\
			\texttt{enum\_tuple\_variant\_field\_missing} & $4$ & $2$ & $2$ \\
			\texttt{tuple\_struct\_to\_plain\_struct} & $4$ & $2$ & $2$ \\
			\texttt{enum\_tuple\_variant\_field\_added} & $3$ & $3$ & $3$ \\
			\texttt{enum\_repr\_int\_removed} & $1$ & $1$ & $1$ \\
			\texttt{enum\_struct\_variant\_field\_added} & $1$ & $1$ & $1$ \Bspace \\
		\hline
	\end{longtable}
\end{center}

All of the detected semver violations, together with their witnesses, can be found in our
GitHub repository \cite{github-csv-results}. They are provided in both human-readable
and machine-readable form. The remaining results (`cargo-semver-checks` false-positive reports
and bugs) are provided only in human-readable form.

\section{The process of gathering results (semver-crater)}\label{r:section_results_semver_crater}

Because cargo-semver-checks is able to create a~rustdoc JSON file
of a~particular crate and release from the registry,
we thought of running the tool on a~large number of existing crates
to find semver violations and bugs such as false-positives or crashes.
We developed a~tool for this purpose and called it \texttt{semver-crater},
analogously to \texttt{crater} which is used by the Rust team to check
new compiler versions on a~large number of crates.

It fetches a~list of one thousand crates available on crates.io sorted by the number of
all-time downloads, which often are the biggest crates and Rust libraries
that strictly adhere to semantic versioning. For each crate
it then retrieves the available releases which were not yanked
and compares every adjacent pair. Lastly it parses the
output of those pairs. The crates were documented only with their
default features, as it avoids instability and better represents the crates'
API which most users would use.
The results are saved to a~CSV file.
Because of the sheer number of releases that needs to be checked, the results take several days
to generate.

Unfortunately, cargo-semver-checks is not yet perfect and sometimes still produces false-positives
that are to be fixed. However, most of them were caused by one of the two cases:
\begin{itemize}
	\item an~item was marked with the \texttt{\#[doc(hidden)]} attribute,
		which hides it from the documentation and thus is not part
		of the public API accessible by the user
		(\texttt{\#[doc(hidden)]} is most often used for code called from inside macro
		implementations as it is not meant to be used directly by the user),
	\item an~item was moved to another crate and re-exported
		(thus the item is still accessible with the old path,
		but its definition is in another rustdoc JSON file making it not visible
		to cargo-semver-checks).
\end{itemize}
Other false-positives also happened, but they showed up rarely and primarily
were just caused by a~bug in the lint query.

Because fixing the two main false-positives turned out to be difficult, we instead shifted
our focus to finding ways for detecting when one of them happened.
We came up with two solutions:
\begin{itemize}
	\item To detect most of the \say{doc-hidden} false-positives
		(though not all), we managed to get a~list of items that are definitely
		\say{doc-hidden} and then ignored such items in the results.
		We were able to create the list by running cargo-semver-checks on one rustdoc JSON file
		containing the hidden items and another one without the hidden items,
		parsing the output to know which items were removed.
	\item As a~way to know whether a~reported item is a~false-positive or a~true-positive,
		we created a~witness (more on that in section \ref{r:section_generating_witnesses})
		and ignored all items whose witnesses compile on both baseline and current releases.
\end{itemize}

Those two techniques greatly reduced the number of reports.
To be sure that the remaining items were true-positives,
we manually checked the code by first generating links to the line of code where each item
was defined and then reading the file.
Mostly we verified whether an~item was indeed a~part of the public API
(not \say{doc-hidden} and not a~part of a~test module).
With that, we have high confidence that the found items are indeed true-positives.

\section{Generating witnesses}\label{r:section_generating_witnesses}

The fact that a~non-major release broke semver can be proven by a~\say{witness},
which consists of two small crates:
\begin{itemize}
	\item a~\say{baseline} -- a~crate which uses the old release as a~dependency
		and compiles fine,
	\item a~\say{current} -- a~crate which uses the new release as a~dependency
		and does not compile.
\end{itemize}
Both crates must have exactly the same code (same \texttt{src/lib.rs} file),
but slightly different manifests.
Thus, a~witness is made of just three short files -- a~source code that uses the item that
broke semver and two manifests that only differ in the \texttt{[dependency]} section
where they have the same crate name as a~dependency, but two different releases of it.

The output of each lint in cargo-semver-checks contains a~list of importable paths
of items that it was triggered by, which is enough to construct many of
the witnesses automatically.

For example, after parsing the output of the \texttt{enum\_variant\_missing} lint,
we got the importable path of the enum and the name of its variant that was removed.
Thanks to that, we were able to write a~Rust code with a~function that deconstructs the enum
to the given variant. This source code compiles in the \say{baseline}, but does not compile
in the \say{current} (because the Rust compiler cannot find the variant), thus it is
a~valid witness.

Using this logic, we wrote a~Python script that parses the output of cargo-semver-checks,
generates the witnesses and tries to compile them.
There are three possible outcomes:
\begin{itemize}
	\item the \say{baseline} compiles and the \say{current} does not compile, which means
		the witness has confirmed that there is a~semver violation (a~true-positive),
	\item both the \say{baseline} and \say{current} compile, which indicates a~false-positive
		or a~bug in one of the dependencies used by cargo-semver-checks,
	\item the \say{baseline} does not compile, which means that the witness
		was constructed incorrectly.
\end{itemize}

We stumbled upon some complications while creating the witnesses for some of the lints.
We had the importable path of an~item, but we were missing the information
about generics or traits, which the Rust compiler requires to be in the source code
(e.g. a~function could take a~generic argument which type needs to be declared when calling
the function). Most cases were solved by writing a~script that iterates over the number of
generic arguments and checks whether a~witness with some number of generic argument
successfully compiles. The other cases mainly required the generic types to satisfy some traits,
which is much harder to automate or even write manually.
Thus we decided that we would skip checking some of the reports that required using generic types.
Because of that, there are 151 witnesses that we did not write
and the releases without the witnesses are \textit{not} included in the statistics in section
\ref{r:section_results_statistics} -- we could have found more semver violations, but it was
too time-consuming for us.

For some lints we could choose between simply importing the given item
(e.g. by \texttt{use importable\_path;}) or creating a~source code that uses it.
Just importing the item can be seen as not convincing and not visually appealing (it does not use
any item properties, e.g. it uses the function name but does not call the function itself).
Thus, we decided to generate witnesses with the more convincing technique of writing
the more realistic Rust code whenever possible and used the pure imports as a~fallback when
it failed to compile.

We had one last difficulty when we tried to create witnesses for
the \texttt{enum\_variant\_added} lint. To construct a~witness for it, one has to exhaustively
pattern-match on all of the given enum's variants. However, the output only contained the name
of the variant that was added rather than a~list of all the variants' names.
To retrieve the full list, we wrote a~separate
Rust project that uses \texttt{trustfall\_rustdoc} -- given the generated rustdoc
of the dependency, it runs a~Trustfall query that finds the enum with matching importable path
and returns all of its variants. Having passed the newly-acquired data to the Python script,
we could then automatically construct its witness.

Some lints were triggered only a~couple of times, so instead of expanding the Python script's
functionality, it was faster to manually write witnesses for them.
There was also the \texttt{struct\_repr\_c\_removed} lint that reported a~breaking change.
However, the issue was based on creating an~undefined behavior rather than a~compilation error,
making it difficult to create a~witness.
Thus we decided to make an~exception and manually verify whether it was indeed a~semver-break.

In the end, our Python script is able to create the following witnesses:
\begin{center}
	\begin{longtable}{| p{6.35cm} | p{8.2cm} |}
		\hline
			\multicolumn{1}{|Sc|}{\textbf{lint name}} &
			\multicolumn{1}{Sc|}{\textbf{witness description}} \\
		\hline
			\begin{tabular}[c]{@{}l@{}}
				\texttt{enum\_missing} \\
				\texttt{struct\_missing}
			\end{tabular} &
			\multicolumn{1}{Sl|}{\parbox{8.2cm}{
				a~function (possibly a~generic one) that takes the missing item as an~argument
			}} \\
		\hline
			\begin{tabular}[c]{@{}l@{}}
				\texttt{trait\_missing}
			\end{tabular} &
			\multicolumn{1}{Sl|}{\parbox{8.2cm}{
				a~generic function (with one or more generic arguments) that takes
				a~generic argument satisfying the missing trait
			}} \\
		\hline
			\begin{tabular}[c]{@{}l@{}}
				\texttt{function\_missing} \\
				\texttt{method\_missing} \\
				\texttt{function\_parameter\_count\_changed} \\
				\texttt{method\_parameter\_count\_changed}
			\end{tabular} &
			\multicolumn{1}{Sl|}{\parbox{8.2cm}{
				a~missing function/method call with an~adequate number of \texttt{todo!()}
				arguments. This technique creates valid witnesses, even though it generates
				warnings that such statements are unreachable (it is still possible to replace
				the \texttt{todo!()} arguments with manually constructed ones, but it is not needed
				to prove a~semver break)
			}} \\
		\hline
			\begin{tabular}[c]{@{}l@{}}
				\texttt{auto\_trait\_impl\_removed} \\
				\texttt{derive\_trait\_impl\_removed}
			\end{tabular} &
			\multicolumn{1}{Sl|}{\parbox{8.2cm}{
				a~function (possibly a~generic one) that takes the item causing a~break
				and another function to which this item is passed by the first function.
			}} \\
		\hline
			\begin{tabular}[c]{@{}l@{}}
				\texttt{enum\_variant\_missing}
			\end{tabular} &
			\multicolumn{1}{Sl|}{\parbox{8.2cm}{
				a~deconstruction of the enum with an~\texttt{if} that requires the missing variant
			}} \\
		\hline
			\begin{tabular}[c]{@{}l@{}}
				\texttt{enum\_variant\_added} \\
				\texttt{enum\_marked\_non\_exhaustive} \\
				\texttt{struct\_field\_added}
			\end{tabular} &
			\multicolumn{1}{Sl|}{\parbox{8.2cm}{
				pattern-matching on all of the item's variants or fields
			}} \\
		\hline
	\end{longtable}
\end{center}

\chapter{Team}\label{r:chapter_team}

\section{Used methodology}\label{r:section_used_methodology}

The project's functionalities and goals can be split into independent improvements, hence our
contributions were mostly composed of multiple smaller changes. The usual workflow of writing
code was as follows:
\begin{enumerate}
	\item discussing with our mentor (who is also the project maintainer) about which improvements
		have the highest priority,
	\item individually picking a~few tasks based on our preferences,
	\item implementing those tasks (and sometimes also thinking about the design of the change)
		and creating Pull Requests on the project's GitHub repositories,
	\item going through an~extensive review process with our mentor (as there are often other
		contributors in the project besides us, the code's readability is a~strong focus point
		in the project),
	\item getting the Pull Request merged by our mentor and releasing (usually a~few days later)
		a~new version of the binary to the registry.
\end{enumerate}

There were some exceptions to this workflow:
\begin{itemize}
	\item at the beginning, we split into two pairs and our mentor gave each of us introductory
		tasks to work on together and familiarize with different aspects of the existing code,
	\item to speedup the process of merging the Pull Requests, we sometimes reviewed each others'
		code before passing it to our mentor,
	\item when a~user reported a~bug, we often looked into the problem and searched for a~temporary
		fix of the bug as soon as possible,
	\item when we encountered bugs or thought of possible additional improvements in the tool's
		functionality, we created an~appropriate Issue on the GitHub repository.
\end{itemize}

Additionally, we held weekly meetings with the four of us to discuss the progress and future
contributions (we did not need to have regular meeting with our mentor, because we had constant
and immediate contact with him through chat).

\section{Responsibilities}\label{r:section_responsibilities}

\begin{itemize}
	\item Tomasz \cite{responsibilities-tomasz}:
		\begin{itemize}
			\item Reported a~bug with an~incorrect path to the generated rustdoc when passing
				a~manifest path in the CLI.
			\item Changed how the manifest of the baseline is built and added selecting all
				features to baseline's manifest, which resolved an~issue reported by a~user of
				the tool where the current and baseline's rustdoc was generated with different
				features, which resulted in false-positives.
			\item Reported a~bug where the tests failed on the newest version of the compiler
				and implemented a~fix.
			\item Renamed directories with queries and tests, thanks to which the structure of
				the project is now clear, more intuitive and updated \texttt{CONTRIBUTING.md}.
			\item Replaced code duplication with a~Rust macro and made adding new lints easier.
			\item Improved the tests, which sped up the process of adding new lints.
				More precisely:
				\begin{itemize}
					\item split each test into separate pairs of crates (that are baseline
						and current crates for the tool),
					\item made it so that the tool runs each lint on all testing crates
						(which has detected multiple new false-positives),
					\item massively improved the readability and ease of adding new tests
						and lints,
					\item improved the error messages when the test output did not match.
				\end{itemize}
			\item Made a~change in the tests that runs the tool on crates that did not change
				to detect possible false-positives.
			\item Committed changes suggested by a~new version of the Rust formatter.
			\item Found and fixed a~bug where the printed baseline version was incorrect.
			\item Added a~new type of integration tests, which locally checks how the tool behaves
				with given command-line arguments, and implemented a~test that checks whether
				a~recent bug (where the lint files were not present in the binary) was fixed.
			\item Diagnosed a~problem where the tests stopped working, fixed the adapter after
				a~new rustdoc version which stopped exporting inlined items of external crates
				and created an~issue for strengthening back a~lint that was affected
				by those changes.
			\item Reported a~problem where the user could be confused by the lack of detailed
				information in the help messages of the tool.
			\item Reported a~problem where the output of the tool on the \texttt{bevy-audio} crate
				seemed like it was producing a~false positive.
			\item Reported a~bug where a~lint's output was duplicated on the
				\texttt{bevy-render} crate.
			\item Added a~feature to disable the usage of the vendored \texttt{OpenSSL} because
				there were problems with it amongst users having the Arch Linux
				desktop distribution.
			\item Changed the way in which the rustdoc of a~crate is generated -- instead of
				directly running the rustdoc command on a~local copy of the crate, it is now ran
				on a~temporary placeholder project, which just has the checked crate as
				a~dependency. It solves a~bug encountered by a~user where running the tool updates
				the \texttt{Cargo.lock} file, and allows to easily specify the set of
				crate's features to run the rustdoc command on.
			\item Reported a~bug where the tool failed to work when generating rustdoc of a~crate
				that only has yanked releases.
			\item Reported a~bug where the tool prints some internal data while checking
				the \texttt{clap} crate.
			\item Found and fixed a~bug where the CI job wrongly checked the exit code in one
				of the integration tests.
			\item Added a~lint in the CI for checking the rustdoc of the project.
			\item Fixed a~broken link which the users saw in the tool's output.
			\item Reported a~false-positive which was related to re-exporting types, values
				and macros with the same names.
			\item Fixed a~user-encountered bug where implicit features due to target-specific
				optional dependencies were not added in the current version.
			\item Fixed a~user-reported bug where the rustdoc command's output was not showing up
				with colors.
			\item Diagnosed a~user-encountered compilation error which happened because of changes
				in one of the tool's dependencies.
			\item Took part in the development of the library based on the existing tool's code.
			\item Developed \texttt{semver-crater} that runs cargo-semver-checks on the most
				popular crates (see chapter \ref{r:chapter_semver_crater})
				and wrote numerous scripts for analyzing the results, designated for:
				\begin{itemize}
					\item parsing the output of cargo-semver-checks to get the list of
						affected items,
					\item filtering releases which were affected by the \say{doc-hidden}
						false-positives,
					\item creating witnesses for all kinds of triggered lints,
					\item verifying whether the witnesses found true-positives,
					\item getting a~list of items for construction of witnesses
						for lints that check the exhaustiveness of pattern matching.
				\end{itemize}
				Manually verified the results with the rest of the team.
		\end{itemize}

	\item Michał \cite{responsibilities-michal}:
		\begin{itemize}
			\item Added a~new type of tests in the CI -- a~regression test that is checking
				the tool against a~specific version of a~user's crate, to be certain that now
				the tool works and has no false-positives on it.
			\item Added a~test that checks whether a~field in the lints is correctly initialized.
			\item Added an~integration test in the CI for specifying the package name.
			\item Made a~new release workflow in the CI that triggers when publishing a~new version
				of the tool and uploads the new binaries to GitHub releases, thanks to which
				the GitHub Actions (both provided by us and written by users for their
				own purposes) that were using the tool can now just download the binary
				(instead of compiling it), which makes the CI job much faster.
			\item Added a~new lint (trustfall query, test crates, tests) to detect tuple structs
				that changed to plain structs.
			\item Added a~new lint (trustfall query, test crates, tests) that checks whether
				a~trait becomes or stops being unsafe and added an~\texttt{unsafe} type property
				to \texttt{Trait} in the schema and the adapter.
			\item Added a~\texttt{fields} edge to the \texttt{Variant} interface in the schema
				and the adapter.
                        \item Changed the way cargo-semver-checks handles features:
                        \begin{itemize}
                            \item Added CLI flags to specify features a~crate is built with for
								the checks.
                            \item Reworked the code that handles crate configuration and adjusted
                                    the default feature configuration.
                                    Previously, the crates were always built with all
									the features enabled. Currently, the default behavior is
									to enable a~heuristically-chosen subset of features,
									as described in \ref{r:section_cli}
                            \item Implemented all the logic for calculating and building the crate
                                    with the final feature set.
                            \item Developed a~test suite for the feature configurations.
                        \end{itemize}
		\end{itemize}

	\item Tomasz and Michał:
		\begin{itemize}
			\item Added a~new lint (Trustfall query, test crates, tests) to detect field removals
				from a~struct variant in enums and added a~field in the adapter for the lint.
			\item Added a~new lint (Trustfall query, test crates, tests) to detect when function
				and method parameter count changes and added the retrieval of function parameter
				names in the adapter.
			\item Fixed a~bug where the rustdoc of the current and the baseline crates
				were overwriting each other, which has been reported by a~user.
		\end{itemize}

	\item Mieszko \cite{responsibilities-mieszko}:
		\begin{itemize}
			\item Reported an~incorrect behavior of the CLI which was choosing the latest release
				from the registry as a~baseline version even when it had been yanked.
				Introduced a~fix and created a~set of unit tests for choosing the baseline crate.
			\item Reported a~problem in release \texttt{v0.18.2} where the CLI was not looking for
				the baseline in the registry as it was supposed to do.
			\item Reported a~bug where the pre-compiled tool crashed on Windows machines.
				Added a~CI job that checks for similar issues.
			\item Found and fixed a~bug that made the CLI ignore
				the \texttt{-{}-release-type} option.
			\item Developed a~new version of \texttt{cargo-semver-checks-action},
				which consisted of:
			\begin{itemize}
				\item creating a~test repository \cite{responsibilities-mieszko-action-tests}
					and using it to add a~CI workflow that tests the basic functionalities of
					the action, then further developing the CI tests to cover all options added in
					the new version,
				\item making use of \texttt{cargo-semver-checks} built-in baseline choosing logic,
				\item adding support for running the action on Windows- and MacOs-based runners,
				\item introducing several new inputs \texttt{package}, \texttt{exclude},
					\texttt{manifest-path}, \texttt{verbose}, \texttt{release-type}
					and \texttt{rust-toolchain}, which are parsed and passed to the CLI,
				\item using pre-built binaries to speed up the action,
				\item adding baseline rustdoc caching between action runs along with new
					inputs \texttt{shared-key} and \texttt{prefix-key}, which allow the user
					to adjust the caching strategy,
				\item setting up environmental variables for best performance of cargo (disabling
					incremental compilation, enabling colors in the output and sparse
					checkout protocol when possible),
				\item rewriting the documentation of the action.
			\end{itemize}
		\end{itemize}

	\item Bartosz \cite{responsibilities-bartosz}:
		\begin{itemize}
			\item Added several new lints (Trustfall queries, test crates, tests) to detect
				when different items have been marked with the \texttt{must\_use} attribute.
				Writing separate lints for individual items allowed to provide more detailed
				messages for the user, improved the code structure by creating smaller,
				more specialized files and made the final review process easier. The items that
				undergo those checks are:
				\begin{itemize}
					\item enums,
					\item structs,
					\item traits,
					\item functions,
					\item methods,
					\item inherent methods (reported only when one was both moved to a~public trait
						and marked with the attribute).
				\end{itemize}
		\end{itemize}

	\item Mieszko and Bartosz:
		\begin{itemize}
			\item Added a~new lint (Trustfall query, test crates, tests) to detect when
				an~exhaustive public enum has the \texttt{non\_exhaustive} attribute added.
			\item Designed and implemented a~new adapter schema for handling attributes.
				Updated all lints that check attributes to use the new schema, as well as added
				multiple new test cases to the queries.
		\end{itemize}
\end{itemize}

\chapter{Conclusion}\label{r:chapter_conclusion}

After several months of constant development, the tool has found recognition among the developers
as it became faster, easier to add to one's project and gained additional lints preventing
even more semver breaks. Numerous new features together with stability and performance improvements
have been introduced frequently as well to further increase its popularity.

During the research that we conducted in chapter \ref{r:chapter_semver_crater},
we have exposed that semver in Rust is hard to adhere to. Even among the top 1000 most downloaded
crates more than \mbox{1 in 6 of them} has violated semver and approximately
\mbox{1 in 30 releases} contained at least one semver violation. Those top crates were built and
are maintained by highly experienced Rust developers, but this is not the case for less popular
ones -- with their developers having less expertise and poorer semver knowledge, they are even more
vulnerable to cause breaks.

So far, cargo-semver-checks has earned the trust of many projects' maintainers. Among them are some
of the largest and most iconic Rust crates, such as tokio (library for concurrent programming)
and libp2p (modular library for peer-to-peer networking). There are also numerous projects from
companies widely known outside the Rust community that adopted the tool as well, including Amazon,
Microsoft, Adobe and Mozilla. Developers have even built third party integrations to bring
cargo-semver-checks into other systems -- the tool is available in nixOS and has been introduced
into release managers like release-plz. With those and many more usage scenarios,
cargo-semver-checks has proven its versatility in the vast, diverse world of semantic versioning.

There are multiple reasons behind cargo-semver-checks great popularity. While it
is impossible to point out a~single, main factor, those are the several (with no specific order)
that resulted in the tool's undoubtful success:
\begin{itemize}
	\item regular implementation of new lints that increased the tool's effectiveness,
	\item introduction of the pre-built executable binaries with their ease of use,
	\item long-term stability thanks to its unique and innovative concept of using Trustfall
		to parse rustdoc,
	\item release and improvements of the GitHub action for continuous integration,
	\item numerous optimizations of the tool's performance (done both by us in
		cargo-semver-checks and by our mentor in the Trustfall engine),
	\item staying in touch with the tool's users and addressing their requests and suggestions,
	\item regular tweets and blog posts by our mentor, giving the community insight into our work
		as well as announcements of upcoming features before new releases.
\end{itemize}

Sometimes the feedback we received from the tool's users was not just limited to a~single comment
-- there were cases where after our response they would share some useful
knowledge or even help with implementing the features they initially asked for. A great example
comes from a~user who was just wondering whether we could do something to make the tool run faster
in continuous integration \cite{make-ci-runs-faster}. Shortly after learning more about our active
goals regarding performance at that time, he decided to do some useful research himself
and provided us with its helpful results. What is more, he was so excited about the possible
improvements that he decided to contribute and implemented them \cite{user-contribution-1}.

The final impact of our work has been an~order of magnitude greater than we expected.
Since the early days of our contributions, we could see the constant increase
in cargo-semver-checks usage as it was growing to its current, well-established position in
the Rust ecosystem. Given its wide and enthusiastic adoption by both corporate and hobbyist
projects, and the high frequency of accidental semantic versioning violations when it is not
in use, the tool's ultimate goal of becoming a~built-in feature of cargo seems closer than it has
ever been before.

Moreover, it is not only the packages' maintainers, but also all of the dependents who benefit
from the tool's presence. Thus its actual impact and the group of beneficiaries are even greater
than the numbers may suggest. With that, we firmly believe that its future is bright and are happy
to already see it making such an~impact, hoping to soon allow everyone to be fearless about running
\texttt{cargo update} on their projects.


\appendix

\end{document}